\documentclass[10pt, conference, compsocconf]{IEEEtran}
\usepackage{graphicx}
\usepackage{amsmath,amssymb}
\usepackage{bm}
\usepackage[algo2e]{algorithm2e}
\usepackage{algorithm,algpseudocode}
\usepackage{algorithmicx}
\usepackage{float}
\ifCLASSINFOpdf
\else
\fi
\begin{document}
%
\title{Vehicle Speed Aware Computing Task Offloading and Resource Allocation Based on Multi-Agent Reinforcement Learning in a Vehicular Edge Computing Network}



%

\author{Xinyu Huang, Lijun He*, Wanyue Zhang\\
\textit{School of Information and Communications Engineering} \\ 
\textit{Xi'an Jiaotong University, Xi'an 710049, China}\\
\textit{{xinyu\_huang@stu.xjtu.edu.cn}, lijunhe@mail.xjtu.edu.cn, Baymaxyue@163.com}}


\maketitle

\begin{abstract}
For in-vehicle application, the vehicles with different speeds have different delay requirements. However, vehicle speeds have not been extensively explored, which may cause mismatching between vehicle speed and its allocated computation and wireless resource. In this paper, we propose a vehicle speed aware task offloading and resource allocation strategy, to decrease the energy cost of executing tasks without exceeding the delay constraint. First, we establish the vehicle speed aware delay constraint model based on different speeds and task types. Then, the delay and energy cost of task execution in VEC server and local terminal are calculated. Next, we formulate a joint optimization of task offloading and resource allocation to minimize vehicles' energy cost subject to delay constraints. MADDPG method is employed to obtain offloading and resource allocation strategy. Simulation results show that our algorithm can achieve superior performance on energy cost and task completion delay.

\end{abstract}


%
\IEEEpeerreviewmaketitle

\section{Introduction}
With the rapid development of Internet of Things, the smart in-vehicle applications (i.e., autonomous driving, image assisted navigation and multimedia entertainment) have been widely applied to smart vehicles \cite{Feng}\cite{Zhang}, which can provide more comfortable and safer environment for drivers and passengers. Since these in-vehicle applications will consume huge computation resource and require low execution latency, the cloud server is employed to afford these complex computing tasks, which results in serious burden on backhaul network \cite{Huang0}. Luckily, vehicular edge computing (VEC) with powerful computing capacity is deployed in the roadside densely and attached in the road side unit (RSU). Therefore, it is worthy to study how to efficiently utilize the computing capacity of VEC server to support low latency and energy-efficiency in-vehicle service. 

To improve the offloading efficiency of vehicle terminals, the researchers have proposed many offloading and resource allocation methods in the VEC-based network. Zhang et al. \cite{Mao} proposed a hierarchical cloud-based VEC ofﬂoading framework to reduce the execution delay, where a back server was utilized to offer extra computation resource for the VEC server. However, the priority of tasks were not considered in the offloading process, which may cause that delay sensitive tasks (i.e., assisted imaged navigation) cannot be processed timely for high-speed vehicles. To further reduce the task execution delay, Liu et al. \cite{Liu} studied the task offloading problem by treating the vehicles and the RSUs as the two sides of a matching problem to minimize the task execution delay. But the computation capacity of VEC server should be fully exploited to decrease the energy consumption of in-vehicle applications. In addition, Li et al. \cite{Li} considered the influence of time-varying channel on the task offloading strategies and formulated the problem of joint radio and computation resource allocation. However, the variety of vehicle speeds was not considered, which may cause that the high-speed vehicles cannot obtain the sufficient computation and wireless resource. To accelerate the radio and computation allocation process, Dai et al. \cite{Dai} proposed a low-complexity algorithm to jointly optimize server selection, offloading ratio and computation resource, but the task handover between VEC servers may result in the increase in delay and occurrence of lost packets. To make a more accurate offloading decision, Sun et al. \cite{Sun} proposed the task offloading algorithm by determining where the tasks were performed and the execution order of the tasks on the VEC server, but the proposed heuristic algorithm may not obtain a satisfying result when the vehicle's channel state fluctuated frequently. Considering the fluctuation of channel state, Zhan et al. \cite{Zhan} proposed a deep reinforcement learning-based offloading algorithm to make the task offloading decision, which can be applied in the dynamic environment caused by vehicle’s mobility. 

In general, there are still some problems to be solved for the computing task offloading and resource allocation for the in-vehicle applications: (1) The impact of vehicle speed on task delay constraints have been neglected, which results in mismatching between vehicle speed and its allocated computation and wireless resource and cannot guarantee the delay requirements. (2) The vast fluctuation of vehicle's channel state caused by fast mobility has not been considered, which may result in offloading failure. (3) The computation capacity of VEC server has not been fully exploited because of the inaccurate resource allocation strategy. Inspired by the above work, we propose a vehicle speed aware computing task offloading and resource allocation strategy based on multi-agent reinforcement learning. Our work is novel in the following aspects.

\textbf{(1)Vehicle speed aware delay constraint model:} Different types of tasks and vehicle speeds demand various delay requirements for in-vehicle applications. Therefore, we fully analyze the internal relationship among vehicle speed, task type and delay requirements to propose a vehicle speed aware delay constraint model, which can make the task offloading and resource allocation process more accurately. 

\textbf{(2)The calculation of energy consumption and delay for different types of tasks:} Based on the bandwidth and delay requirements, in-vehicle computing tasks are classified into three types: critic application, high-priority application and low-priority application. For different types of tasks, we calculate the energy consumption and delay based on task transmission and execution process for different offloading positions, respectively.

\textbf{(3)Multi-agent reinforcement learning based solution to our formulated problem:} A joint optimization of offloading and resource allocation problem is formulated by a Markov decision process (MDP) process, with the objective to minimize the energy consumption subject to delay constraint. In addition, multi-agent reinforcement learning is applied to solve the high-dimension action decision-making.

\section{System Model}
\subsection{System Framework}
\begin{figure}[!h]
	\centering
	\includegraphics[width=8.5cm]{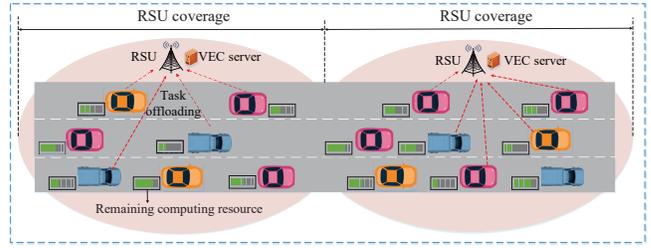}
	\centering
	\caption{The architecture of computing task offloading in VEC network}
	\label{system}
\end{figure}

The scenario considered in this paper is the computing task offloading of vehicles on urban roads in VEC network, as shown in Figure \ref{system}. RSU is located along the roadside, and the coverage areas of adjacent RSUs do not overlap. Therefore, according to the coverage areas of RSUs, the road can be divided into several adjacent segments, where the vehicle can only establish the wireless link with the RSU of the current road segment. Each RSU is equipped with a VEC server whose powerful computation capacity can help the vehicle quickly handle computing tasks. Since the delay constraint of the computing task is extremely short, it can be assumed that the vehicle can still receive the task processing result from the previous VEC server when the vehicles travels to the road segment horizon. In addition, the computing task can also be executed locally to alleviate the traffic and computing burden of VEC server.

Because the VEC server can real-time perceive the state information of vehicles and owns powerful computing processing capacity, so it can generate the optimal computing task offloading and resource allocation strategy (including computing resource allocation and wireless resource allocation) for each vehicle. First, the state information of the vehicle, including task queue, speed, location, remaining computing capacity and wireless resources, is reported to the RSU in real time. The RSU will forward the received state information to the VEC server, which utilizes the extracted state information to calculates the task execution delay and energy consumption to formulate the optimized problem. The goal is to reduce the energy consumption of all vehicles by optimizing task offloading and resource allocation decisions without exceeding delay constraint. Then, according to the results of the joint task offloading and resource allocation, the vehicle's computing tasks can be executed locally or offloaded to the VEC server.

\subsection{Vehicle Speed Aware Delay Constraint Model}

Vehicle' computing tasks can be divided into three types: critical application (CA), high-priority application (HPA) and low-priority application (LPA), which needs different bandwidth and delay requirements \cite{Dzi}. We denote these three task types by $\phi_1$, $\phi_2$ and $\phi_3$, respectively. CA task generally refers to the autonomous driving and road safety tasks, which needs ultra delay to ensure the safety of vehicle driving. Therefore, this type of task needs to be executed locally and delay threshold is set to $Th{{r}_{1}}$. For the HPA task, it mainly involves image assisted navigation, parking navigation and some optional security applications. The delay tolerance for HPA task is related to the current vehicle speed. For vehicles with low speed, it doesn't matter even if the computing task takes a little longer time. More wireless and computation resource can be allocated to vehicles with high speed, whose tasks can be processed preferentially. The delay threshold of HPA task is set to $Th{{r}_{2}}$ when the vehicle speed reaches at the maximum road speed limit ${{v}_{\max }}$. LPA tasks generally includes multimedia and passenger entertainment activities, so the requirement for delay threshold is relatively slack. The delay threshold is set to $Th{{r}_{3}}$.

In this paper, it is assumed that the generated computing tasks are independent task sequence. The computing task of the vehicle $k$ at time $t$ to be processed is defined as ${{\mathcal{I}}_{t}}(k)$. For HPA task, when the vehicle's speed is low, the delay threshold of the computing task can be relatively longer. With the increase of speed, the information of the vehicle received from surrounding environment will increase rapidly at the same delay because of the longer distance traveled. Therefore, the delay of the computing task that the vehicle needs to be processed should be reduced rapidly. When the speed reaches a higher level, with the increase of speed, the increase amplitude of the vehicle's information received from the surrounding environment will gradually decrease. Therefore, the delay of the computing task that the vehicle needs to be processed will be reduced slowly.

Therefore, we select one-tailed normal function to describe the relationship between delay constraint, $\mathcal{T}({{v}_{\mathcal{I}_t(k)}})$, and speed for task $\mathcal{I}_t(k)$ of $\phi _2$, as follows:
\begin{equation}\small
\begin{aligned}
\mathcal{T}({{v}_{\mathcal{I}_t(k)}})&=Th{{r}_{2}}\frac{1}{\sqrt{2\pi }\alpha }\exp (-\frac{v_{{{\mathcal{I}}_{t}}(k)}^{2}}{2{{\alpha }^{2}}})/(\frac{1}{\sqrt{2\pi }\alpha }\exp (-\frac{v_{\max }^{2}}{2{{\alpha }^{2}}})) \\ 
& =\exp (-\frac{v_{{{\mathcal{I}}_{t}}(k)}^{2}-v_{\max }^{2}}{2{{\alpha }^{2}}})Th{{r}_{2}},\text{ }if\text{  }{{\mathcal{I}}_{t}}\text{(}k\text{)}\in \phi _2 \\ 
\end{aligned}
\end{equation}
where ${{v}_{\mathcal{I}_t(k)}}$ is the current vehicle speed and ${{\alpha }^{2}}$ is the variance of the normal function. $v_{max}$ is the road maximum speed. To ensure that the probability that vehicle speed is within the maximum speed exceeds 95\%, we denote $\alpha ={{v}_{\max }}/1.96$. Therefore, We employ $\Upsilon({{v}_{\mathcal{I}_t(k)}})$ to represent the delay threshold of computing task ${{\mathcal{I}}_{t}}(k)$ of all task types, as follows:
\begin{equation}
\begin{aligned}
& \Upsilon ({{\mathcal{I}}_{t}}\text{(}k\text{)})=\left\{ \begin{aligned}
& Th{{r}_{1}},\text{ }if\text{  }{{\mathcal{I}}_{t}}\text{(}k\text{)}\in {{\phi }_{1}} \\ 
& \mathcal{T}\text{(}{{v}_{{{\mathcal{I}}_{t}}\text{(}k\text{)}}}\text{), }if\text{  }{{\mathcal{I}}_{t}}\text{(}k\text{)}\in {{\phi }_{2}} \\ 
& Th{{r}_{3}},\text{ }if\text{  }{{\mathcal{I}}_{t}}\text{(}k\text{)}\in {{\phi }_{3}} \\ 
\end{aligned} \right. \\ 
& \text{           } \\ 
\end{aligned}
\end{equation}

\section{Delay and Energy Consumption of Different Offloading Positions}
\begin{figure}[!h]
	\centering
	\includegraphics[width=8.5cm]{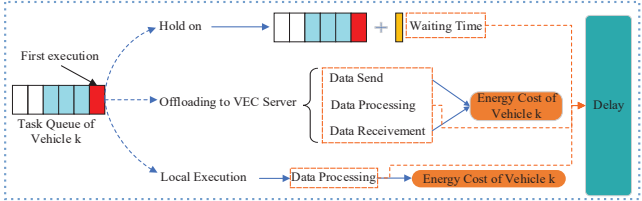}
	\centering
	\caption{Delay and energy consumption of different offloading pcositions}
	\label{delay}
\end{figure}

For the generated computing task, the task handover between VEC server is generally not considered \cite{Zhang} to ensure its successful transmission. For HPA and LPA computing tasks generated by the in-vehicle applications at time $t$, there are usually three ways to handle them, which are hold on, offloading to VEC server and local execution, as shown in Figure \ref{delay}.
When the vehicle's current remaining computing resource and wireless resource and VEC server's remaining computing resource are insufficient to process the new computing task, the vehicle's computing task can choose to wait for a certain time until computing and wireless resources are released. 
\subsection{Delay of task execution}
\subsubsection{Offloading to the VEC server}
For the local VEC server, we denote the set of vehicles in the service area by $\mathbb{Q}$ and the number of these vehicles is $K$. When the computing task ${{\mathcal{I}}_{t}}(k)$ belonging to task type $\phi_2$ or $\phi_3$ is offloaded to VEC server, the task completion time contains upload time, execution time and download time. For the task type $\phi_2$, since the size of output file is much smaller than that of input file, the download time can be ignored. Considering that task upload, execution and download cannot be executed simultaneously within one transport time interval (TTI), the consumed time, $TR_{{{\mathcal{I}}_{t}}(k)}$, needs to round up, which can be described as: 
\begin{equation}\small
TR_{{{\mathcal{I}}_{t}}(k)}^{{}}=\left\{ \begin{aligned}
& \left\lceil \frac{{{c}_{{{\mathcal{I}}_{t}}(k)}}}{r_{k}^{VEC,up}} \right\rceil +\left\lceil \frac{{{\kappa }_{{{\mathcal{I}}_{t}}(k)}}{{c}_{{{\mathcal{I}}_{t}}(k)}}}{b_{{{\mathcal{I}}_{t}}(k)}^{VEC}{{f}^{VEC}}} \right\rceil ,\text{  }if\text{ }{{\mathcal{I}}_{t}}(k)\in \phi_2 \\ 
& \left\lceil \frac{{{c}_{{{\mathcal{I}}_{t}}(k)}}}{r_{k}^{VEC,up}} \right\rceil +\left\lceil \frac{{{\kappa }_{{{\mathcal{I}}_{t}}(k)}}{{c}_{{{\mathcal{I}}_{t}}(k)}}}{b_{{{\mathcal{I}}_{t}}(k)}^{VEC}{{f}^{VEC}}} \right\rceil +\left\lceil \frac{{{\omega }_{{{\mathcal{I}}_{t}}(k)}}{{c}_{{{\mathcal{I}}_{t}}(k)}}}{r_{k}^{VEC,down}} \right\rceil ,\\&\text{ }if\text{ }{{\mathcal{I}}_{t}}(k)\in \phi_3 \\ 
\end{aligned} \right.
\end{equation}
where ${{c}_{{{\mathcal{I}}_{t}}(k)}}$ is the file size of task ${{\mathcal{I}}_{t}}(k)$ and ${{\kappa }_{{{\mathcal{I}}_{t}}(k)}}$ is the calculation density of processing the task ${{\mathcal{I}}_{t}}(k)$. ${{\omega }_{{{\mathcal{I}}_{t}}(k)}}$ is the scaling ratio of the downloaded task size relative to the uploaded task size. $b_{{{\mathcal{I}}_{t}}(k)}^{VEC}$ indicates the proportion of computing resource allocated by VEC server to the task ${{\mathcal{I}}_{t}}(k)$. ${{f}^{VEC}}$ denotes the CPU frequency of VEC server. The transmission capacity between vehicle and server can be obtained from the number of allocated channels, channel bandwidth, transmission power and noise power \cite{Huang}. For uplink channel $n$ of VEC server allocated to vehicle $k$, the available uplink transmission capacity, $r_{k,n}^{VEC,up}$, can be expressed as
\begin{equation}\small
r_{k,n}^{VEC,up}=\omega _{VEC}^{up}{{\log }_{2}}\left( 1+\frac{P\cdot h_{k,n}^{VEC,up}}{{{\sigma }^{2}}+I_{k,n}^{VEC,up}} \right),\text{  }for\text{  }n\in \mathbb{N}_{VEC}^{up}
\end{equation}
where $\omega _{VEC}^{up}=B_{VEC}^{up}/N_{VEC}^{up}$. $B_{VEC}^{up}$ is the uplink bandwidth of VEC server and $N_{VEC}^{up}$ is the number of total channels of VEC server. $\sigma^2$ denotes noise power and $P$ is transmission power. $I_{k,n}^{VEC,up}$ denotes the interference on channel $n$. $\mathbb{N}_{VEC}^{up}$ indicates the uplink channel set of VEC server. Let $z_{k,n}^{VEC,up}$ indicate whether the uplink channel $n$ is allocated to the vehicle $k$. If it is allocated, $z_{k,n}^{VEC,up} = 1$, otherwise, $z_{k,n}^{VEC,up} = 0$. Then the uplink transmission capacity between vehicle $k$ and VEC server, $r_{k}^{VEC,up}$, can be depicted as
\begin{equation}
r_{k}^{VEC,up}=\sum\limits_{n\in \mathbb{N}_{VEC}^{up}}{z_{k,n}^{VEC,up}r_{k,n}^{VEC,up}}
\end{equation}

Similarly, the transmission capacity of downlink channel $n$ of VEC server allocated to vehicle $k$, $r_{k,n}^{VEC,down}$, can be expressed as
\begin{equation}
\begin{aligned}
r_{k,n}^{VEC,down}&=\omega _{VEC}^{down}{{\log }_{2}}\left( 1+\frac{P\cdot h_{k,n}^{VEC,down}}{{{\sigma }^{2}}+I_{k,n}^{VEC,down}} \right),\\& \text{  }for\text{  }n\in \mathbb{N}_{VEC}^{down}
\end{aligned}
\end{equation}
where $\omega _{VEC}^{down}=B_{VEC}^{down}/N_{VEC}^{down}$. Then the downlink transmission capacity between vehicle $k$ and VEC server, $r_{k}^{VEC,down}$, can be depicted as
\begin{equation}
r_{k}^{VEC,down}=\sum\limits_{n\in {{\mathbb{N}_{VEC}^{down}}}}{z_{k,n}^{VEC,down}r_{k,n}^{VEC,down}}
\end{equation}

\subsubsection{Local Execution}
For computing task ${{\mathcal{I}}_{t}}(k)$ executed locally, the consumed time, $TL_{{{\mathcal{I}}_{t}}(k)}$, can be expressed as:
\begin{equation}
TL_{{{\mathcal{I}}_{t}}(k)}^{{}}=\left\lceil \frac{{{\kappa }_{{{\mathcal{I}}_{t}}(k)}}{{c}_{{{\mathcal{I}}_{t}}(k)}}}{{{f}^{k}}} \right\rceil 
\end{equation}
where ${{f}^{k}}$ is the CPU frequency of vehicle $k$. At time $t$, computing task ${{\mathcal{I}}_{t}}(k)$ can select to hold on, be offloaded to the local VEC server and be executed locally. Therefore, for computing task ${{\mathcal{I}}_{t}}(k)$, the total delay from generation to completion, $D({{\mathcal{I}}_{t}}(k))$, is derived by
\begin{equation}\small
\begin{aligned}
D({{\mathcal{I}}_{t}}(k))&=t-t_{{{\mathcal{I}}_{t}}(k)}^{g}+\tau _{{{\mathcal{I}}_{t}}(k)}^{Hold}{{T}_{h}}\\&+(1-\tau _{{{\mathcal{I}}_{t}}(k)}^{Hold})[\tau _{{{\mathcal{I}}_{t}}(k)}^{VEC}TR_{{{\mathcal{I}}_{t}}(k)}^{{}}+(1-\tau _{{{\mathcal{I}}_{t}}(k)}^{VEC})T{{L}_{{{\mathcal{I}}_{t}}(k)}}]
\end{aligned}
\end{equation}
where $t_{k,i}^{g}$ is the generated time of task ${{\mathcal{I}}_{t}}(k)$.  $\tau _{{{\mathcal{I}}_{t}}(k)}^{Hold}$ indicates whether task ${{\mathcal{I}}_{t}}(k)$ holds on. If it holds on, ${{\mathcal{I}}_{t}}(k) = 1$, otherwise, ${{\mathcal{I}}_{t}}(k) = 0$. ${{T}_{h}}$ denotes the waiting time. $\tau _{{{\mathcal{I}}_{t}}(k)}^{VEC}$ indicates whether task ${{\mathcal{I}}_{t}}(k)$ is offloaded to VEC server. If it is allocated, $\tau _{{{\mathcal{I}}_{t}}(k)}^{VEC} = 1$, otherwise, $\tau _{{{\mathcal{I}}_{t}}(k)}^{VEC} = 0$.

\subsection{Energy Consumption of Task Execution}
\subsubsection{Offloading to VEC server}
When the computing task is offloaded to the VEC server, the energy consumption origins from uploading and downloading computing task. For the task type $\phi_2$, since the size of output file is much smaller than that of input file, the energy consumed by downloading computing task can be ignored. Therefore, the energy consumption of task $\mathcal{I}_{t}(k)$ belonging to $\phi_2$ or $\phi_3$ offloaded to VEC server, $ER_{{{\mathcal{I}}_{t}}(k)}$, can be depicted as
\begin{equation}
ER_{{{\mathcal{I}}_{t}}(k)}^{{}}=\left\{ \begin{aligned}
& P\frac{{{c}_{{{\mathcal{I}}_{t}}(k)}}}{r_{k}^{VEC,up}},\text{  }if\text{  }{{\mathcal{I}}_{t}}(k)\in \phi_2 \\ 
& P(\frac{{{c}_{{{\mathcal{I}}_{t}}(k)}}}{r_{k}^{VEC,up}}\text{+}\frac{{{\omega }_{{{\mathcal{I}}_{t}}(k)}}{{c}_{{{\mathcal{I}}_{t}}(k)}}}{r_{k}^{VEC,down}}),\text{  }if\text{  }{{\mathcal{I}}_{t}}(k)\in \phi_3 \\ 
\end{aligned} \right.
\end{equation}

\subsubsection{Local Execution}
When computing task ${{\mathcal{I}}_{t}}(k)$ is executed locally, the consumed energy can be calculated according to the assigned computation resource, $EL_{{{\mathcal{I}}_{t}}(k)}$, which can be expressed as
\begin{equation}
EL_{{{\mathcal{I}}_{t}}(k)}^{{}}={{\xi }_{{{\mathcal{I}}_{t}}(k)}}{{\kappa }_{{{\mathcal{I}}_{t}}(k)}}{{c}_{{{\mathcal{I}}_{t}}(k)}}{{({{f}^{k}})}^{2}}
\end{equation}
where ${\xi }_{\mathcal{I}_{t}(k)}$ is the energy density of processing task ${{\mathcal{I}}_{t}}(k)$ \cite{Din}.
According to the different offloading positions of computing task ${{\mathcal{I}}_{t}}(k)$, including VEC server and local device, the consumed energy of all vehicles served by local VEC server, $E(t)$, can be derived by
\begin{equation}
E(t)=\sum\limits_{k\in \mathbb{Q}}{(1-\tau _{{{\mathcal{I}}_{t}}(k)}^{Hold})[\tau _{{{\mathcal{I}}_{t}}(k)}^{VEC}E{{R}_{{{\mathcal{I}}_{t}}(k)}}}+(1-\tau _{{{\mathcal{I}}_{t}}(k)}^{VEC})E{{L}_{{{\mathcal{I}}_{t}}(k)}}]
\end{equation}

\section{Delay and Energy-Efficiency Driven Computing Task Offloading and Resource Allocation Algorithm Based on Multi-Agent Reinforcement Learning}
\subsection{Problem Formulation}
We formulate the optimized problem of reducing the energy consumption of each vehicle without exceeding delay constraint by carrying out optimal computing task offloading and resource allocation strategy, which can be described as follows:
\begin{equation}\label{problem}
\begin{aligned}
& \underset{{{X}_{t}}(k,n),\forall k,n}{\mathop{\min }}\,\sum\limits_{t=1}^{T}{E(t)} \\ 
& s.t. \\ 
& (c1)\tau _{{{\mathcal{I}}_{t}}(k)}^{VEC}+\tau _{{{\mathcal{I}}_{t}}(k)}^{Hold}\le 1,\forall k\in \mathbb{Q} \\ 
& (c2)\sum\limits_{k}{b_{{{\mathcal{I}}_{t}}(k)}^{VEC}}\le 1,\forall k\in \mathbb{Q} \\ 
& (c3)\sum\limits_{k}{z_{k,n}^{VEC,up}}\le 1,\forall k\in \mathbb{Q},n\in \mathbb{N}_{VEC}^{up} \\ 
& (c4)\sum\limits_{k}{z_{k,n}^{VEC,down}}\le 1,\forall k\in \mathbb{Q},n\in \mathbb{N}_{VEC}^{down} \\ 
& (c5)D({{\mathcal{I}}_{t(k)}})\le \Upsilon({{v}_{\mathcal{I}_t(k)}}),\forall k\in \mathbb{Q} \\ 
\end{aligned}
\end{equation}
where ${{X}_{t}}(k,n)=(\tau _{{{\mathcal{I}}_{t}}(k)}^{VEC},\tau _{{{\mathcal{I}}_{t}}(k)}^{Hold},z_{k,n}^{VEC,up},z_{k,n}^{VEC,down})$. Constraint (c1) implies that computing task ${{\mathcal{I}}_{t(k)}}$ cannot be offloaded to the local VEC server, executed locally and hold on simultaneously. Constraint(c2) indicates that computation capacity allocated to computing task ${{\mathcal{I}}_{t(k)}}$ by VEC server cannot exceed its own computing capacity. Constraint (c3) and constraint (c4) indicate that each channel must be assigned to one and only one vehicle at each scheduling period. Constraint (c5) indicates that computing task ${{\mathcal{I}}_{t(k)}}$ should be completed within the delay constraint.

\subsection{Deep Reinforcement Learning-Based Solution Method}
Equation (\ref{problem}) is a multi-vehicle cooperation and competition problem, which is obviously a NP-hard problem. Therefore, we employ the deep reinforcement learning method to solve the proposed computing task offloading and resource allocation problem. First, we formulate our problem as a MDP to accurately describe the offloading and resource allocation decision process and utilize the multi-agent deep deterministic policy gradient (MADDPG) \cite{Lowe} to find the optimal policy for the MDP. In what follows, we will present the elements of MDP, including state space, action space and reward function.
\subsubsection{State Space}
We defined the state space of vehicle $k$ as ${{s}_{k}}(t)$, including the state information of vehicle $k$, other vehicles and VEC server, which is depicted as
\begin{equation}\small
\begin{aligned}
{{s}_{k}}(t)&=[{{v}_{1}}(t),...,{{v}_{K}}(t),{{d}_{1}}(t),...,{{d}_{K}}(t),{{c}_{1}}(t),...,{{c}_{K}}(t), \\&r{{b}_{VEC}}(t),
s\tau _{k}^{Hold}(t),s\tau _{k}^{VEC}(t),sb_{k}^{VEC}(t), sz_{1}^{VEC,up}(t),\\&...,sz_{N_{VEC}^{up}}^{VEC,up}(t),sz_{1}^{VEC,down}(t),...,sz_{N_{VEC}^{down}}^{VEC,down}(t)] \\ 
\end{aligned}
\end{equation}
where ${{v}_{k}}(t),{{d}_{k}}(t),{{c}_{k}}(t)$ are vehicle speed, position and file size to be processed of vehicle $k$ at time $t$, respectively. $r{{b}_{VEC}}(t)$ is the current remaining computation capacity of VEC server at time $t$. $s\tau _{k}^{(\centerdot )}(t)$ indicates whether the vehicle $k$ selects the offloading position $(\centerdot)$ at time $t$. If it is selected, $s\tau _{k}^{(\centerdot )}(t) = 1$, otherwise,  $s\tau _{k}^{(\centerdot )}(t)$ = 0. $sb_{k}^{VEC}(t)$is the ratio of computation resource allocated by VEC server to vehicle $k$ at time $t$. $sz_{1}^{VEC,up}(t),...,sz_{N_{VEC}^{up}}^{VEC,up}(t)$ indicates whether the uplink channel resource of VEC is available at time $t$. If it is available, the value is 1, otherwise, the value is 0. $sz_{1}^{VEC,down},...,sz_{N_{VEC}^{down}}^{VEC,down}(t)$ indicates whether the downlink channel resource of VEC is available at time $t$. If it is available, the value is 1, otherwise, the value is 0. Therefore, the state space of the system can be defined as: ${{S}_{t}}=({{s}_{1}}(t),...{{s}_{k}}(t)...,{{s}_{K}}(t))$.

\subsubsection{Action Space}
Since vehicle $k$ cannot offload multiple tasks simultaneously at time $t$, which means that task ${{\mathcal{I}}_{t(k)}}$ and vehicle $k$ have a one-to-one correspondence. Therefore, We can represent the action space of task ${{\mathcal{I}}_{t(k)}}$ with the one of vehicle $k$. For vehicle $k$, the action space, $a_k(t)$, contains whether to hold on and offload task to VEC server, computation resource allocated by VEC server and the uplink and downlink channels allocated by VEC server, which can be expressed as
\begin{equation}\small
\begin{aligned}
a_{k}^{{}}(t)&=[\tau _{k}^{Hold}(t),\tau _{k}^{VEC}(t),b_{k}^{VEC}(t),z_{k,1}^{VEC,up}(t),...,z_{k,N_{VEC}^{up}}^{VEC,up}(t),\\&z_{k,1}^{VEC,down}(t),...,z_{k,N_{VEC}^{down}}^{VEC,down}(t)]
\end{aligned}
\end{equation}
Therefore, the action space of the system can be defined as: ${{A}_{t}}=\{{{a}_{1}}(t),...{{a}_{k}}(t)...,{{a}_{K}}(t)\}$.

\subsubsection{Reward}
The goal of this paper is to reduce the energy consumption of each vehicle terminal without exceeding task delay constraint, which can be realized by allocating the computation resource and wireless resource of the system. Therefore, we set rewards based on constraint conditions and objective function to accelerate the training speed. After taking action ${{a}_{k}}(t)$, if the state of vehicle $k$ does not satisfy the constraints (c1)-(c4), the reward function can be defined as 
\begin{equation}\small
\begin{aligned}
& {{r}_{k}}(t)={{\ell }_{1}}+{{\Gamma }_{1}}\cdot (s\tau _{k}^{VEC}+s\tau _{k}^{Hold}-1)\cdot {{\Lambda }_{(s\tau _{k}^{VEC}+s\tau _{k}^{Hold}\le 1)}}\\&+{{\Gamma }_{2}}\cdot (\sum\limits_{k}{sb_{k}^{VEC}}-1)\cdot {{\Lambda }_{(\sum\limits_{k}{sb_{k}^{VEC}}\le 1)}}\\& 
+{{\Gamma }_{3}}\cdot (\sum\limits_{k}{sz_{k,n}^{VEC,up}}-1)\cdot {{\Lambda }_{(\sum\limits_{k}{sz_{k,n}^{VEC,up}}\le 1)}}\\&+{{\Gamma }_{4}}\cdot (\sum\limits_{k}{sz_{k,n}^{VEC,down}}-1)\cdot {{\Lambda }_{(\sum\limits_{k}{sz_{k,n}^{VEC,down}}\le 1)}} \\ 
\end{aligned}
\end{equation}
where ${{\Lambda }_{(\centerdot )}}$ indicates that if the condition $(\centerdot)$ is not satisfied, the value is -1, otherwise, the value is 0. ${{\ell }_{1}},{{\Gamma }_{1}},{{\Gamma }_{2}},{{\Gamma }_{3}},{{\Gamma }_{4}}$ is experimental parameters. After taking action ${{a}_{k}}(t)$, if the state of vehicle $k$ satisfy all constraints (c1)-(c4), the reward function can be defined as 
\begin{equation}
{{r}_{k}}(t)={{\ell }_{2}}+\exp (Th{{r}_{k}}(t)-\Upsilon({{v}_{\mathcal{I}_t(k)}}))
\end{equation}
where ${{\ell }_{2}}$ is experimental parameters. After taking action ${{a}_{k}}(t)$, if the state of vehicle $k$ satisfy all constraints (c1)-(c5), the reward function can be defined as 
\begin{equation}
r(t)={{\ell }_{3}}\text{+}{{\Gamma }_{5}}\cdot \exp ({{E}_{k}}(t))
\end{equation}
where ${{\ell }_{3}},{{\Gamma }_{5}}$ denote experimental parameters.
\subsubsection{Joint Delay and Energy-Efficiency Algorithm Based on MADDPG}
The centralized training process is composed of $K$ agents, whose network parameter are $\theta =\{{{\theta }_{1}},...,{{\theta }_{K}}\}$. We denote $\mu =\{{{\mu }_{{{\theta }_{1}}}},...,{{\mu }_{{{\theta }_{K}}}}\}$ (abbreviated as $\mu_i$) as the set of all agent deterministic policies. So for the deterministic policy ${{\mu }_{k}}$ of agent $k$, the gradient can be depicted as
\begin{equation}
\begin{aligned}
&{{\nabla }_{{{\theta }_{k}}}}J({{\mu }_{k}})=\\&{{\mathbb{E}}_{S,A\sim \mathcal{D}}}[{{\nabla }_{{{\theta }_{k}}}}{{\mu }_{k}}({{a}_{k}}|{{s}_{k}}){{\nabla }_{{{a}_{k}}}}Q_{k}^{\mu }(S,{{a}_{1}},...,{{a}_{K}}){{|}_{{{a}_{k}}={{\mu }_{k}}({{s}_{k}})}}]
\end{aligned}
\end{equation}
where $\mathcal{D}$ is experience replay buffer, which contains a series of $(S,A,{{S}^{'}},R)$. $Q_{k}^{\mu }(s,{{a}_{1}},...,{{a}_{K}})$ is the Q-value function. For the critic network, it can be updated according to the loss function as follows
\begin{equation}
\begin{aligned}
& \mathcal{L}({{\theta }_{k}})={{\mathbb{E}}_{S,A,R,{{S}^{'}}}}{{[Q_{k}^{\mu }(S,a_{1}^{{}},...,a_{K}^{{}})-y)}^{2}}] \\ 
& \text{where} \;\; y=r_{k}^{{}}+\gamma Q_{k}^{{{\mu }^{'}}}({{S}^{'}},a_{1}^{'},...,a_{K}^{'}){{|}_{a_{j}^{'}=\mu _{j}^{'}({{s}_{j}})}} \\ 
\end{aligned}
\end{equation}
where $\gamma $ is the discount factor. The action network is updated by minimizing the policy gradient of the agent, which can be expressed as
\begin{equation}
{{\nabla }_{{{\theta }_{k}}}}J\approx \frac{1}{X}\sum\limits_{j}{{{\nabla }_{{{\theta }_{k}}}}}{{\mu }_{k}}(s_{k}^{j}){{\nabla }_{{{a}_{k}}}}Q_{k}^{\mu }({{S}^{j}},a_{1}^{j},...,a_{K}^{j}){{|}_{{{a}_{k}}={{\mu }_{k}}(s_{k}^{j})}}
\end{equation}
where $X$ is the size of mini-batch, $j$ is the index of samples. The specific joint delay and energy-efficiency algorithm based on MADDPG (JDEE-MADDPG) is shown in Algorithm 1.
\begin{algorithm}[!h]
	\caption{JDEE-MADDPG}
	\label{alg1}
	\textbf{Initialize:} the positions, speed, task queue, computing resources and wireless resources of all vehicles. Initialize the computing and wireless resources of the VEC server. Initialize the weights of actor and critic networks.
	
	\For{{episode= 1:M}}
	{
		Initialize a random process $\mathcal{N}$ for action exploration;
	
		Receive initial state $S$;
		
		\For{each vehicle $k=1,...,K$}
		{
			Execute actions ${{a}_{k}}$ and obtain new state $s_{k}^{'}$;
			
			\uIf{ the $s_{k}^{'}$ does not satisfy constraints (c1)-(c4) in Eq.(12):}
				{
				 	Obtain the reward of vehicle $k$ based on Eq.(15);
			    }
			\uElseIf{the $s_{k}^{'}$ satisfy all constraints (c1)-(c4) in Eq.(12):}	 
				{
					Obtain the reward of vehicle $k$ based on Eq.(16);
				}
			\uElseIf{the $s_{k}^{'}$ satisfy all constraints (c1)-(c5) in Eq.(12):}
				{
					Obtain the reward of vehicle $k$ based on Eq.(17);
				}

			 \textbf{end}
			 
             Obtain	the action $A$, new state ${{S}^{'}}$ and reward $R$;
             
             Store $(S,A,{{S}^{'}},R)$ in replay buffer $\mathcal{D}$;
              
		}
		\For{each vehicle $k=1,...,K$}
		{
			Sample a random mini-batch of $X$ samples $({{S}^{j}},{{A}^{j}},{{R}^{j}},{{S}^{'}}^{j})$ from $\mathcal{D}$;
			
			Update the critic network by minimizing the loss function, Eq.(19);
			
			Update actor network using the sampled policy gradient, Eq.(20);
		} 
		Update the target network parameters of each vehicle $k$: $\theta _{k}^{'}\leftarrow \delta {{\theta }_{k}}+(1-\delta )\theta _{k}^{'}$
	}
\end{algorithm}

\section{Simulation Results }
\subsection{Parameter Setting}
The specific simulation parameters are presented in Table I and Table II. The algorithms compared in this section are as follows: 

\textbf{All Local Execution (AL):} All computation tasks are executed locally.

\textbf{All VEC Execution (AV):} The CA tasks are executed locally, while HPA and LPA tasks are executed in VEC server. The resource allocation strategy is based on the size of task.

\textbf{Random Offloading (RD):} The HPA and LPA tasks are executed locally and in VEC server based on the uniform distribution. The resource allocation strategy is based on the size of task.

\textbf{Energy and Delay Greedy (EDG):} The offloading strategy is based on vehicle’s channel state and resource allocation strategy is based on the size of task, in order to decrease the energy cost and execution delay in each step. 

\begin{table}[!h]
	\caption{Simulation Parameter Configuration}
	\begin{tabular}{|c|c|ccc}
		\cline{1-2}
		\bf{Parameter}                                                                  & \bf{Value}                                &  &  &  \\ \cline{1-2}
		Number of vehicles                                                                    & 5, 7 ,9, 11, 13                               &  &  &  \\ \cline{1-2}
		Size of   task queue                                                                  & 10                                            &  &  &  \\ \cline{1-2}
		Size of task   input                                                                  & {[}0.2, 1{]} Mb                               &  &  &  \\ \cline{1-2}
		Speed of vehicle                                                                      & {[}30, 50{]}, {[}50, 80{]}, {[}30, 80{]} Km/h &  &  &  \\ \cline{1-2}
		RSU’s   coverage range                                                                & 500 m                                         &  &  &  \\ \cline{1-2}
		RSU’s   bandwidth                                                                     & 100 MHz                                       &  &  &  \\ \cline{1-2}
		Channel model                                                                         & Typical Urban                                 &  &  &  \\ \cline{1-2}
		\begin{tabular}[c]{@{}c@{}}Transmission power between\\  vehicle and RSU\end{tabular} & 0.5 W                                         &  &  &  \\ \cline{1-2}
		\begin{tabular}[c]{@{}c@{}}Computation   capacity \\ of VEC server\end{tabular}       & 10 G Cycles/s                                   &  &  &  \\ \cline{1-2}
		\begin{tabular}[c]{@{}c@{}}Computation   capacity \\ of vehicle\end{tabular}          & 1, 1.2, 1.4, 1.6, 1.8 G Cycles/s                &  &  &  \\ \cline{1-2}
		Computation   density                                                                 & {[}20, 50{]} Cycles/bit                       &  &  &  \\ \cline{1-2}
		Waiting   time of hold on                                                             & 20, 50 ms                                     &  &  &  \\ \cline{1-2}
		Delay Threshold                                 & 10, 40, 100ms                                     &  &  &  \\ \cline{1-2}		
		\begin{tabular}[c]{@{}c@{}}Output data   size/ input \\ data size ratio\end{tabular}  & 0.1                                           &  &  &  \\ \cline{1-2}
		Energy density {\cite{Dai}}                                             & $1.25\times 10^{-26}$ J/Cycle                            &  &  &  \\ \cline{1-2}
		Parameters of reward                                                                  &  \begin{tabular}[c]{@{}c@{}}${{\Gamma }_{1}}=0.8$, ${{\Gamma }_{2}},{{\Gamma }_{3}},{{\Gamma }_{4}},{{\Gamma }_{5}}=0.5$ \\  ${{\ell }_{1}}=-0.4,{{\ell }_{2}}=-0.2,{{\ell }_{3}}=0.5$ \end{tabular}                                       &  &  &  \\ \cline{1-2}
	\end{tabular}
\end{table}

\begin{table}[!h]
	\caption{The Neural Network and Training Parameters}
	\begin{tabular}{|c|c|c|c|}
		\hline
		\textbf{Parameter}                 & \textbf{Value} & \textbf{Parameter}        & \textbf{Value}  \\ \hline
		Layers                             & 3              & Layer Type                & Fully Connected \\ \hline
		\multicolumn{1}{|c|}{Hidden Units} & 512            & Learning Rate of   Critic & 0.001           \\ \hline
		Optimizer                          & Adam           & Learning Rate of Actor    & 0.0001          \\ \hline
		Episode                            & 140000         & Activation Function       & Relu            \\ \hline
		Mini-batch                          & 128            & Buffer Size               & 20000           \\ \hline
	\end{tabular}
\end{table}

\subsection{Performance Evaluation}
We validate the algorithm performance in term of convergence property, task completion delay and energy consumption  under different simulation configurations. 

\begin{figure}[!h]
	\centering
	\includegraphics[width=8.6cm]{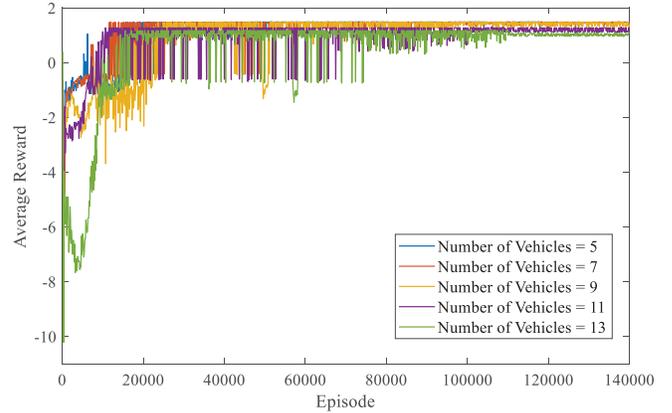}
	\centering
	\caption{Convergence property of different numbers of vehicles.}
	\label{convergence}
\end{figure}
\begin{figure*}[!h]
	\centering
	\includegraphics[width=\textwidth]{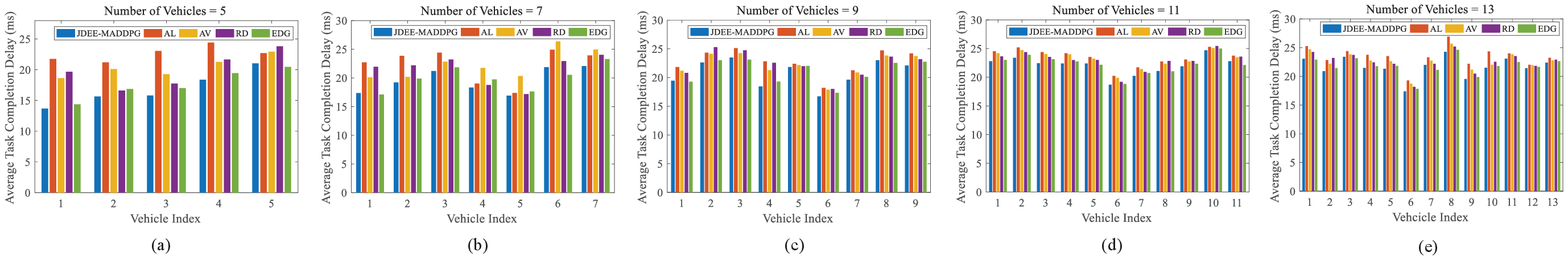}
	\centering
	\caption{Average task completion delay of different algorithms: (a) the number of vehicles is 5, (b) the number of vehicles is 7, (c) the number of vehicles is 9, (d) the number of vehicles is 11, (e) the number of vehicles is 13.}
	\label{figure1}
\end{figure*}
\begin{figure*}[!h]
	\centering
	\includegraphics[width=\textwidth]{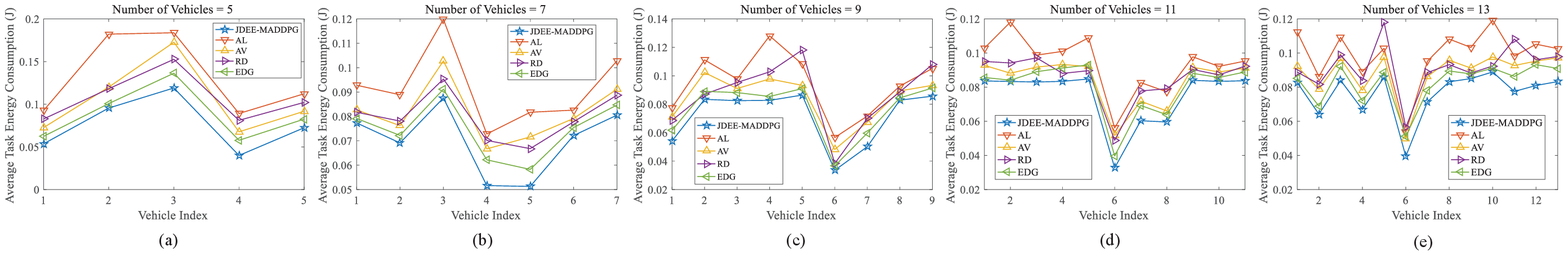}
	\centering
	\caption{Average task energy consumption of different algorithms: (a) the number of vehicles is 5, (b) the number of vehicles is 7, (c) the number of vehicles is 9, (d) the number of vehicles is 11, (e) the number of vehicles is 13.}
	\label{figure2}
\end{figure*}
\begin{figure*}[!h]
	\centering
	\includegraphics[width=\textwidth]{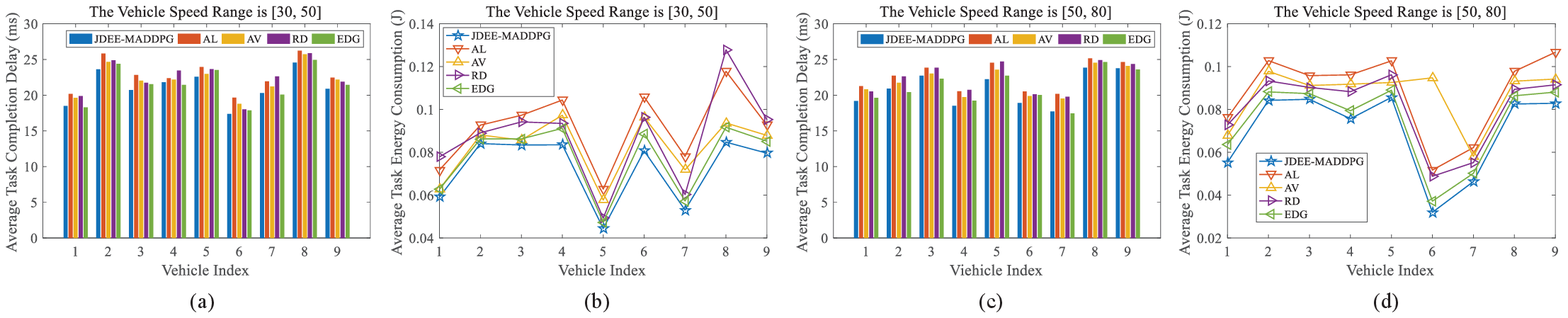}
	\centering
	\caption{Average task completion delay and energy consumption of different algorithms: (a)(b) the vehicle speed range is [30, 50], (c)(d) the vehicle speed range is [50, 80].}
	\label{figure3}
\end{figure*}

In Figure \ref{convergence}, we present the convergence performance of our proposed JDEE-MADDPG algorithm with the different numbers of vehicles. It can be seen that with the increase of training episodes, the average reward of vehicles rises gradually and preserves a stable positive reward eventually. In the initial stage, the average reward of our proposed JDEE-MADDPG algorithm with less vehicles is higher than that with more vehicles, because the increase of vehicles means higher dimension of state space and action space and our proposed JDEE-MADDPG algorithm needs to take more explorations. Therefore, it will result in that the average reward of large number of vehicles is lower than that of small number of vehicles. As the training episode increases, our proposed JDEE-MADDPG algorithm gradually achieves the convergence state, where the average reward of 5 vehicles is the highest, while the average reward of 13 vehicles is the lowest. The reason is that more vehicles implies that they will compete the limited computation resource and wireless resource more fiercely, which causes that the reward of energy consumption and delay decreases. 

In Figure \ref{figure1}, we present the comparison of average task completion delay with different numbers of vehicles when the vehicle speed range is from 30 to 80 Km/h. It can be seen that compared with AL, AV and RD algorithms, our proposed JDEE-MADDPG algorithms can always preserve a lower level of task completion delay for each client. This is because that our proposed JDEE-MADDPG algorithm can allocate the computation resource and wireless resource to vehicles more accurately based on the task priority, task size, vehicle speed and vehicle's channel state. In addition, the task completion delay of some vehicles of EDG algorithm is less than that of our proposed JDEE-MADDPG algorithm, because our proposed algorithm sacrifices a little task completion delay to decrease the energy consumption of vehicle terminals without exceeding the task delay constraint.

In Figure \ref{figure2}, we show the comparison of average task energy consumption with different numbers of vehicles when the vehicle speed range is from 30 to 80 Km/h. It can be observed that compared with other algorithms, our proposed JDEE-MADDPG algorithm can always preserve a lower level of energy consumption. This is because that our proposed JDEE-MADDPG algorithm can always make the optimal offloading and resource allocation strategy based on the task priority, task size, vehicle speed and vehicle’s channel state and reduce the energy consumption of all vehicles as soon as possible.

In Figure \ref{figure3}, we compare the average task completion delay and energy consumption with different vehicle speed range when the number of vehicles is 9. Compared with AL, AV and RD algorithm, our proposed JDEE-MADDPG algorithm performs better in terms of delay and energy consumption. This is because that our proposed JDEE-MADDPG algorithm can utilize more information of vehicle terminals and VEC server, i.e., vehicle position, vehicle speed, task queue, channel state and remaining computation resource to make the optimal offloading and resource allocation strategy. Besides, the reason that some vehicles' task completion delay of our proposed JDEE-MADDPG algorithm remains higher than that of EDG algorithm is that our JDEE-MADDPG algorithm usually allocate more wireless and computation resource of VEC server to the vehicles with high speed without exceeding task delay constraint, which may cause that task completion delay of some vehicles is higher than that of EDG algorithm.

\section{Conclusion}
In this paper, we propose a vehicle speed aware computing task offloading and resource allocation algorithm to achieve the goal of energy-efficiency for all vehicles within task delay constraint. First, we establish the vehicle speed-based delay constraint model based on task types and vehicle speed. And then we calculate the task completion delay and energy consumption for different offloading positions based on the allocated computation and wireless resource. Finally, we formulate the mathematical model with the objective to minimize energy consumption of all vehicles subject to the delay constraint. The MADDPG method is utilized to obtain the offloading and resource allocation strategy. Simulation results show that the proposed JDEE-MADDPG algorithm can decrease energy consumption and task completion delay compare with other algorithms under different numbers of vehicles and vehicle speed ranges.

\section*{Acknowledgement}
This research work was supported in part by the National Science Foundation of China (61701389, U1903213), the Natural Science Basic Research Plan in Shaanxi Province of China (2018JQ6022) and the Shaanxi Key R\&D Program (2018ZDCXL-GY-04-03-02).

\end{document}